\documentclass[sigplan,10pt]{acmart}
\usepackage{graphicx}

\usepackage[ruled, vlined, linesnumbered]{algorithm2e}

\startPage{1}

\setcopyright{none}

\usepackage{booktabs}   
\usepackage{subcaption} 

\begin{document}

\title[Short Title]{GraphVine: A Data Structure to Optimize Dynamic Graph Processing on GPUs}         


\author{Rohith Krishnan S}
\affiliation{
  \department{Department of Computer Science and Engineering}              
  \institution{Indian Institute of Technology Ropar}            
  \city{Rupnagar}
  \state{Punjab}
  \postcode{Post-Code1}
  \country{India}                    
}
\email{2021csm1005@iitrpr.ac.in}          

\author{Venkata Kalyan Tavva}
\affiliation{
  \department{Department of Computer Science and Engineering}              
  \institution{Indian Institute of Technology Ropar}            
  \city{Rupnagar}
  \state{Punjab}
  \postcode{Post-Code1}
  \country{India}                    
}
\email{kalyantv@iitrpr.ac.in}          

\author{Rupesh Nasre}
\affiliation{
  \department{Department of Computer Science and Engineering}              
  \institution{Indian Institute of Technology Madras}            
  \city{Chennai}
  \state{Tamil Nadu}
  \postcode{Post-Code1}
  \country{India}                    
}
\email{rupesh@cse.iitm.ac.in}          


\begin{abstract}
Graph processing on GPUs is gaining momentum due to the high throughputs observed compared to traditional CPUs, attributed to the vast number of processing cores on GPUs that can exploit parallelism in graph analytics. This paper discusses a graph data structure for dynamic graph processing on GPUs. Unlike static graphs, dynamic graphs mutate over their lifetime through vertex and/or edge batch updates. The proposed work aims to provide fast batch updates and graph querying without consuming too much GPU memory. Experimental results show improved initialization timings by 1968-1269024\%, improved batch edge insert timings by 30-30047\%, and improved batch edge delete timings by 50-25262\% while consuming less memory when the batch size is large.
\end{abstract}



\maketitle

\section{Introduction}
\label{intro_sec}
Graph processing is vital to several problems, from forming the basic structure of social, communication and supply-chain networks to solving optimization problems (by mapping them to graph problems). Graph problems are highly parallelizable since multiple threads can work on different graph entities (like edge, vertex, sub-graphs, etc.,) at once, making them ideal for GPU processing. A graph data structure stores the input graph to be processed and forms the backbone of graph processing. The insert, delete and querying (search) performance of such underlying data structures determine the performance of the required graph processing.

The two general classes of graph processing include \textit{static} and \textit{dynamic} graph processing. Static graph processing involves working with input graphs with a fixed initial set of vertices and edges, making management of the graph data  straightforward. Generally, graphs are stored using traditional graph representations like adjacency matrices and lists. However, fitting the graphs within the GPU memory using these methods becomes impossible when the graphs become too large. Hence better graph representation schemes that use less space, like \textit{CSR} (Compressed Sparse Row) \cite{graph}, \textit{CRS} (Compressed Row Storage), and\textit{ CCS} (Compressed Column Storage), are used. These representations reduce the storage requirement to a fraction of what adjacency matrices use. They also provide much better query timings than adjacency lists, making them a good fit for static graph processing. On the other hand, in the case of dynamic graphs, these representations prove costly since the graph data structure needs a rebuild for every single update to the graph. Therefore, it is essential to develop graph data structures for dynamic graph processing that perform proper memory management, and efficient inserts, deletes, and searches on the dynamic graph data structure.

Owing to the complexity associated with GPU processing, there are only a few works in literature on dynamic graph processing on GPUs. Current state-of-the-art works on dynamic graph processing include \textit{GPMA and GPMA+} \cite{gpma}, \textit{diff-CSR} \cite{diffcsr}, \textit{cuSTINGER} \cite{custinger}, \textit{Hornet} \cite{hornet}, \textit{aimGraph} \cite{aimgraph}, \textit{faimGraph} \cite{faimgraph}, and \textit{SlabHash} \cite{slabhash}. Among these works, \textit{faimGraph} and \textit{SlabHash} are the only ones supporting vertex addition and deletion. Remaining works consider a scenario wherein the vertex data remains static and only the edge data mutates. Depending on the input graphs, the performance of each of the above frameworks varies drastically, with each framework performing well only on a subset of graphs while performing badly for other graphs. The related works section below discusses more details regarding each of these techniques.

Our proposed graph data structure provides fast vertex and edge inserts, deletes, and querying on dynamic graphs. The adjacency list is in the form of edge blocks, each holding multiple edges. On demand, a sufficient number of edge blocks gets allocated to each source vertex. After the delete operations, the data structure performs memory reclamation, and the reclaimed memory is used in future batch updates as per the need. Our contributions to this work are:

\begin{itemize}
  \item We propose a graph data structure that performs fast vertex and edge inserts, deletes, and querying.
  \item We propose a novel \textit{edge pre-allocated queue}, where edge blocks are allocated on the GPU memory beforehand. Hence future requests from adjacency lists for further edge blocks are immediately addressed.
\end{itemize}
\section{Prior Work}
\label{PW_sec}
There have only been a handful of dynamic graph data structures. A significant challenge encountered by each was to come up with an adjacency list structure that accommodates future batch updates. 

David Bader et al. \cite{custinger} proposed \textit{cuSTINGER}, that supports dynamic edge inserts and deletes while considering the number of vertices as static. Since the number of vertices is static, it uses a \textit{Structure of Array} (SoA) representation to store edge data associated with each source vertex. Each attribute of this edge data, like weight and last modified time, is stored in multiple arrays. The \textit{Array of Structures} (AoS) representation could cause a considerable overhead on CUDA-capable GPUs regarding memory requests. The structure of array representation offers better locality if multiple threads perform updates on the same source vertex. \textit{cuSTINGER} simplifies the insert and delete procedures by separating them, unlike the predecessors that used to do it concurrently. Edge inserts in \textit{cuSTINGER} involve performing an initial duplicate check within the batch as well as with the existing graph, and then inserted into the adjacency list associated with a vertex (the structure of array representation mentioned above represents the adjacency list of a particular vertex). If the adjacency list gets filled up during an edge update, the entire list gets copied to a larger location, followed by inserts of the new edges. Edge deletes involve locating and marking the edge location as deleted and replacing the edge at the end of the list with the deleted edge. Vertex inserts and deletes get treated as a series of edge inserts and deletes. 

The same team propose a follow-up work of the above \textit{cuSTINGER}, \textit{Hornet} \cite{hornet}, which performs fast inserts and deletes for large update batches, which was the issue with \textit{cuSTINGER}. Like its predecessor, \textit{Hornet} only supports edge insertions and considers the number of vertices static. Each vertex in \textit{Hornet} is associated with two attributes: \textit{the number of neighbours associated with the vertex} and a \textit{pointer to the adjacency list associated with the vertex}. The edge blocks responsible for storing the edge details in \textit{Hornet} have sizes in multiples of 2. Initialization involves locating empty blocks for each vertex’s adjacency list based on its degree. Adjacency lists of all the nodes are initially kept in the host memory. Later all the initialized data is copied to the GPU memory in bulk, maximizing the PCI-Express bandwidth. During inserts, if an edge block is full, it is allocated a new edge block with a size double that of the previous block. This process is done by querying empty blocks from an array of B+ trees (each index is associated with edge blocks of size 2\textsuperscript{index}), where each index in the B+ tree array has pointers to all B+ trees with that particular size. All previous entries get copied to the new edge block, and later using a vectorized bit-tree, the empty location within the edge block gets located, and the new edge is inserted. Edge deletes follow a similar approach, the difference being that if the size-reduced edge block could fit in a smaller edge block, the new block is fetched from the B+ tree array followed by copying the contents of the current block to the new smaller block. There also exists duplicate removals within the update batch and with the graph, as seen in \textit{cuSTINGER}. Vertex inserts and deletes are performed as a series of edge inserts and deletes. 

Martin Winter et al. \cite{aimgraph} propose \textit{aimGraph}, that supports edge inserts and deletes while keeping the number of vertices constant. During initialization, \textit{aimGraph} allocates the entire GPU memory in one memory allocation call, and all subsequent allocation calls made by the internal memory manager of \textit{aimGraph} allocate space for edge blocks from the initially allocated GPU memory. This single memory allocation saves significant time as round-trips to the CPU for \textit{cudaMalloc} calls are no longer necessary, and calls to the internal memory manager are sufficient for additional storage for the edge blocks. Since \textit{aimGraph} does not support vertex inserts or deletes, vertex data is considered static. The memory manager allocates the static data during initialization while preprocessing the input graph during initialization allocates sufficient memory for adjacency lists associated with each vertex. Subsequent edge inserts on future batch updates query for additional space to the memory manager if required. During subsequent deletes, the memory manager does not reclaim empty blocks, assuming they will get used during future insert operations to the same source vertex. 

The same team later proposed \textit{faimGraph} \cite{faimgraph}, which supported edge and vertex inserts and deletes. Like \textit{aimGraph}, \textit{faimGraph} also allocates the entire GPU memory using one single \textit{cudaMalloc} call, and the internal memory manager handles further memory allocations. Vertex data on \textit{faimGraph} grows from the top to bottom, while the edge data from the bottom to top for efficient utilization of available memory space. \textit{faimGraph} also introduces two new data structures: vertex and page queue, where upon deallocating vertex and edge blocks, the pointers to these blocks get added to their respective queues. On subsequent allocation requests, the respective queues get searched first, and if they are not empty, the pointer is passed to the requesting thread, saving a significant amount of time. Since vertex data is dynamic, using a \textit{Structure of Array} (SoA) representation could prove troublesome. As the number of vertices increases or decreases, changes to the arrays within the structure cause a considerable overhead as the size of the vertex deletes and inserts become high. An \textit{Array of Structure} (AoS) representation seems perfect here, as it does not need such changes. Due to the availability of the vertex and edge queues, memory allocation is achieved in \textit{O(1)} time, provided pointers to blocks are available on the respective queues. Vertex inserts comprise a duplicate check followed by a straightforward insert, while vertex deletes involve edge deletes of all edges associated with the vertex, both incoming and outgoing. Compaction of the adjacency lists is done after vertex deletes. Edge inserts involve sorting the edges alongside inserting them into the concerned adjacency list in \textit{O(V)} time, where \textit{V} is the number of edges in the adjacency list of the source vertex. Edge deletes are straightforward; after deleting the edge, it is replaced with the last edge on the adjacency list, followed by moving elements across it to maintain the sorted order. Like \textit{aimGraph}, the graphs taken for evaluation were from the 10th DIMAC Graph Implementation Challenge. Due to the efficient memory reclamation and usage, \textit{faimGraph} significantly improved performance over \textit{aimGraph} and \textit{cuSTINGER}, while also providing a much better memory management technique than \textit{aimGraph}.

Other works on dynamic graphs on GPUs include \textit{Accelerating Dynamic Graph Analytics on GPUs} \cite{gpma} and \textit{Fast Dynamic Graph Algorithms} \cite{diffcsr}. The former discusses a technique called \textit{GPMA} (GPU-based Packed Memory Array) and \textit{GPMA+}, which follow a self-balancing binary tree structure divided into many levels. A locking mechanism across the levels of the self-balancing tree does edge updates. The latter proposes a modified \textit{Compressed Sparse Row} (CSR) format called \textit{diff-CSR} to support vertex and edge inserts and deletes. The subsequent edge inserts during a batch update are reflected only in the \textit{diff-CSR} array and not in the original \textit{CSR} array. \textit{Hornet} and \textit{faimGraph} significantly outperformed both these works, as was the case with \textit{cuSTINGER} and \textit{aimGraph}.

Our proposed work is the first to support dynamic vertex and edge updates while consuming far less memory, unlike \textit{faimGraph}, and helps handle large graphs and additional graph analytic computations.

\section{Motivation}
\label{Motive_sec}
In \textit{Accelerating Dynamic Graph Analytics on GPUs}, not much is said about the memory management unit of \textit{GPMA} and \textit{GPMA+}. It also tends to use a large amount of memory, and a significant effort is taken to reallocate the entries evenly across the self-balanced binary tree. These problems would get amplified when the update batch is enormous. In \textit{Fast Dynamic Graph Algorithms}, the proposed \textit{diff-CSR} mechanism could cause large insert and delete overheads with a large update batch due to the need to rearrange and concatenate the \textit{diff-CSR} throughout multiple updates. \textit{cuSTINGER}, on the other hand, performs individual \textit{cudaMalloc} calls to allocate additional edge blocks to adjacency lists from the CPU, causing significant overheads, especially if the update batch is large. It also incurs huge penalties when relocating adjacency lists if they become full while over-allocating space causing poor efficiency. \textit{Hornet} takes care of many of these issues, but due to the high initialization overheads, it performs relatively poorly than \textit{cuSTINGER} in small update batches. Both \textit{aimGraph} and \textit{faimGraph} allocate the entire GPU memory during initialization, meaning that performing any advanced analytic computation in the GPU is impossible. They are also slower than \textit{Hornet} for large batches since adjacency lists in \textit{Hornet} are stored in one single edge block, while in \textit{aimGraph} and \textit{faimGraph}, it gets stored in multiple edge blocks. \textit{faimGraph}, on the other hand, is much better in terms of memory efficiency than \textit{aimGraph} and \textit{Hornet}, as \textit{aimGraph} does not return an empty edge block in the hope that the source vertex associated with the edge block reuses it again, while \textit{Hornet} tends to over-allocate space. Through this work, we aim to devise a solution that addresses all these issues. In the next section, we present the details of the proposed GraphVine data structure along with the details of its initialization, edge/vertex updates (inserts and deletes) and search operations.

\section{GraphVine structure}
\label{Our_sec}

The proposed work performs fast insert, delete, and query operations regardless of batch size, all while consuming less memory. \textit{Figure 1} gives an overview of the proposed graph data structure. The \textit{vertex dictionary} holds the vertex data and is a single contiguous array. The adjacency list associated with each vertex comprises an \textit{edge sentinel node} that holds metadata regarding the adjacency list. \textit{Edge blocks} hold edge data, and the edge sentinel node has a pointer to the edge blocks. The following sections perform an in-depth look into each component.



\begin{figure}[tbh]
\begin{center}
\includegraphics[width=0.6\columnwidth]{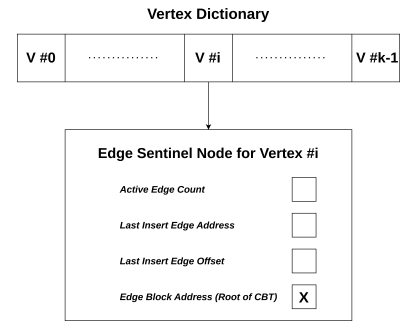} \\
\caption{GraphVine Data Structure}
\label{fig1}
\end{center}
\end{figure}

\subsection{Vertex dictionary and adjacency list structure}
\label{ssec:vd}

As mentioned above, the vertex dictionary is a single contiguous array holding vertex data. Each vertex entry in the vertex dictionary holds the vertex id and a pointer to the adjacency list associated with that vertex. Adding more attributes to each vertex entry would affect spatial locality in the GPU and performance. The size of the vertex dictionary at any stage would be the closest power of 2 to the vertex size. For example, for a graph with 453 vertices, the vertex dictionary size is 512.

Each vertex entry in the vertex dictionary has a pointer to the edge sentinel node of that vertex, and forms the basic adjacency list structure. The edge sentinel node holds the vertex's edge count, the edge block address and the offset within that edge block where the latest insertion happened. Multiple edge blocks hold edge data in the form of a \textit{Complete Binary Tree} (CBT) as in \textit{Figure 2}, and the edge sentinel node holds the address of the root of the CBT. Each edge block follows an \textit{Array of Structures} (AoS) format, each structure denoting an edge entry. The edge entry may have multiple attributes depending on the graph's requirement. For a vertex, the number of edge blocks an adjacency requires depends its degree.

\begin{center}
\begin{figure}
\includegraphics[width=0.6\columnwidth]{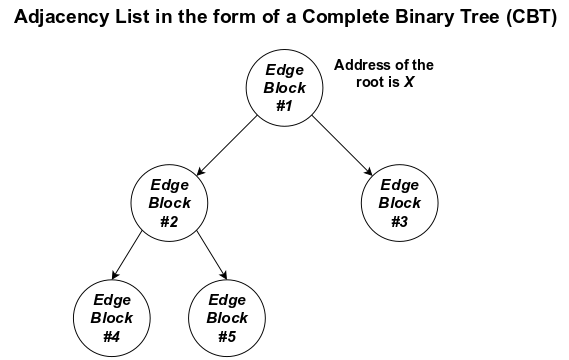} \\
\caption{The proposed CBT of edge blocks. Each edge block holds multiple edge entries, and the address of the root of the CBT (say, \textit{X}) is stored in the edge sentinel node of the vertex, as presented in \textit{Figure 1}.}
\end{figure}
\end{center}

Since the adjacency follows a CBT structure, the tree height for a given set of edge blocks would be the lowest. The tree is not a \textit{Binary Search Tree} (BST), eliminating the need for height-balanced BSTs like \textit{AVL} trees. An AVL tree would be beneficial for search operations with a single thread, but since we have the luxury of multiple thread launches for an adjacency, an unsorted binary tree would work just fine. During an edge update, if all edge blocks in the current adjacency are full, additional edge blocks are allocated from the edge queue per the new requirement of the source vertex. The below section discusses the edge queue in detail.

\subsection{Edge Queue}
\label{ssec:eq}
Due to the mutation of dynamic graphs, the edge blocks required by an adjacency list may change over time. Using \textit{cudaMalloc} calls to allocate edge blocks to an adjacency list on demand is very costly. Instead, we pre-allocate edge blocks and push their addresses onto an edge queue. Adjacency lists requiring additional edge blocks will pop edge block addresses from the queue, cutting the time required for CPU round-trips during a \textit{cudaMalloc} call. In the event of edge deletes, if edge blocks of an adjacency list get empty, their addresses are pushed to the edge queue for future reuse.

Upon launch of the graph data structure, at least half of the GPU memory gets reserved through a single \textit{cudaMalloc} call for use as the edge queue. Once the graph data structure exhausts 80\% of the edge queue, another \textit{cudaMalloc} call pushes 25\% more edge block addresses to the queue if there is sufficient memory. Else, as many possible blocks get pushed.

\subsection{Initialization and batch updates}
\label{ssec:init}
The initialization phase determines several parameters of the graph data structure. The edge block size, the number of edge entries within an edge block, is the average degree of the non-zero degree vertices in the first batch update. Experimentation revealed that adding zero-degree vertices to the equation has a detrimental effect on determining the correct edge block size and performance. As explained in the edge queue section above, half the GPU memory is reserved for the edge blocks using a single \textit{cudaMalloc} call, keeping future batch updates in mind. From the experimental results, these numbers are more than sufficient. Though the edge block size remains constant, pushing additional edge blocks to the edge queue is repeated for each batch update according to the adjacency requirement.

In dynamic graph processing, the graph starts with an initial set of vertices and edges, followed by batch updates that insert or delete vertices and edges from the graph. The proposed graph data structure assumes the batch updates are in a \textit{Compressed Sparse Row} (CSR) format. Otherwise, the proposed work converts the input graph format into \textit{CSR} before commencing inserts or deletes.

\subsection{Dynamic vertex updates}
\label{ssec:dv}
As mentioned before, the proposed graph data structure supports dynamic vertex updates. The size of the vertex dictionary is always a power of two. The initialization of the vertex dictionary on the GPU is as follows. First, one \textit{cudaMalloc} call allocates space on the GPU memory for the vertex dictionary with a capacity equal to the closest power of two of the vertex size of the first update batch. Second, another \textit{cudaMalloc} call allocates space for the edge sentinel nodes. Third, a GPU kernel gets launched with the number of threads equal to the number of vertices in the first batch and assigns one edge sentinel node to each source vertex, completing the basic adjacency list structure. Vertices added to the vertex dictionary in the later batch updates follow the same initialization procedure, where a new edge sentinel node gets allocated.

Later with future batch updates, once the allocated size becomes insufficient to store new vertices, the existing values get copied to a new vertex dictionary twice the previous size, which would again be a power of two. One \textit{cudaMalloc} call allocates the new location for the vertex dictionary, and one GPU kernel call copies the contents of the previous vertex dictionary to the new location. Later, deallocation of the previous vertex dictionary memory takes place. If the batch update performs vertex deletes, the graph data structure does not reclaim memory since the adjacency lists do the majority of the space taken by the data structure.

\subsection{Dynamic edge updates}
\label{ssec:de}
Edge updates are much more common on dynamic graphs than vertex updates. The proposed dynamic graph data structure assumes that the batched edge updates are in a \textit{CSR} format. If not, it gets converted into the corresponding \textit{CSR} format before the batch processing. The update batch then gets copied to the GPU, and multiple threads are launched depending on the adjacency operation, namely inserts and deletes. The edge insert procedure is discussed below.


\subsubsection{Dynamic edge inserts}

Once the update batch in the \textit{CSR} format is ready, the edge inserts commence. For the first batch, the number of edge blocks a source vertex requires is calculated depending on its degree. This data gets pushed into a vector the size of the number of vertices in the graph. Then a prefix sum calculation is done on this vector, followed by a calculation of a space remaining vector at the CPU side, where it keeps track of the number of empty edge entries in the last edge block that did insertions for each source vertex. The proposed batched edge insertion algorithm is \textit{vertex-centric}, hence the GPU kernel responsible for the edge inserts is launched with the number of threads equal to the number of vertices in the graph, meaning that each source vertex gets a thread allocated for the insert operation.

\begin{center}
\begin{figure}
\includegraphics[width=0.6\columnwidth]{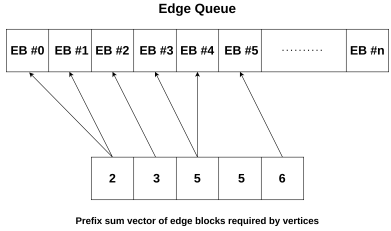} \\
\caption{The graph data structure parallelly popping the edge block addresses from the edge queue with the help of a prefix sum vector of edge blocks required by each source vertex in the graph. The number of entries in the prefix sum vector is equal to vertex count of the input graph.}
\end{figure}
\end{center}

\begin{figure}[tbh]
\begin{center}
\includegraphics[width=0.6\columnwidth]{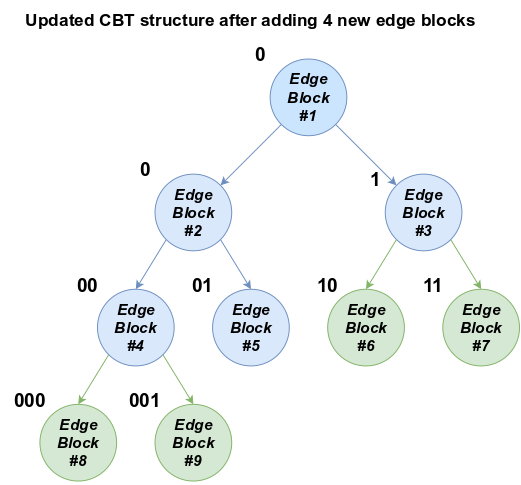} \\
\caption{An example demonstrating how the CBT is updated from the second batch update to an adjacency. Blue and Green edge blocks represent the old and new edge blocks in the CBT, respectively.}
\label{edge_update_123}
\end{center}
\end{figure}

If the degree of a source vertex in the update batch is zero, the associated thread performs nothing. For a source vertex with a non-zero degree, the thread pops edge blocks from the edge queue depending on its requirement. All the launched threads can simultaneously pop edge blocks from this queue in parallel due to the prefix sum vector of the edge block counts of each source vertex as in \textit{Figure 3}. CUDA allows each thread that gets launched to have a distinct identifier value, so each thread needs to check the values in the prefix sum array for its corresponding index to get details of the indices in the edge queue that it needs to pop. For example, a thread with \textit{ID \#1} checks the prefix sum array of edge blocks at indices \textit{0} and \textit{1} to get the number of edge blocks required and the indices in the edge queue that need popping. A thread with \textit{ID\#0} checks the prefix sum vector at index \textit{0} only since it is the first entry in the prefix sum vector. This method allows popping edge blocks in parallel instead of using atomic operations to avoid race conditions. Our initial implementation used CUDA's inbuilt \textit{atomicCAS} (Compare And Swap) operations in the edge queue, and the performance penalty was very high compared to the prefix sum generation and parallel popping mechanism.

Once each thread acquires the edge block addresses it requires, it generates a Complete Binary Tree (CBT) structure with the edge blocks. For the first batch, the CBT generation is straightforward and similar to any generic one. Then the thread updates the edge sentinel node with the address of the root of the CBT. The thread then calculates the start and end index of the edges of the source vertex associated with, with the help of the \textit{CSR} offset and edge fields. Then it iterates through each edge block in the CBT and inserts the edges in sequence, starting from the edge block that forms the root of the CBT. Once an edge block gets full, the edges get inserted into the next edge block in the CBT, which is empty. Since we calculated the edge blocks required by an adjacency in the batch pre-processing stage, we do not run out of edge blocks and have the exact number required for insertions. After inserting the final edge of the adjacency, we update the last insert edge offset and last insert edge block entries in the edge sentinel node. Finally, a GPU kernel gets launched that updates the front pointer of the edge queue.

The edge insert procedure is quite different on the subsequent batch updates to the adjacency of a source vertex. The number of edge blocks required by a source vertex gets calculated considering the \textit{space remaining} in the edge blocks of the adjacency from the previous batch update. The number of threads launched by the GPU kernel for edge inserts remains the same, equal to the number of vertices in the graph. The parallel popping of edge blocks happens as usual, followed by adding the new edge blocks to the CBT structure. Unlike the first batch's CBT generation, inserting the new edge blocks into the CBT is not straightforward. Hence for each new edge block, we generate a \textit{bit string} that decides its location in the existing CBT. For example, in \textit{Figure 4} for the newly generated edge block which is the ninth edge block in the CBT, the bit string generated would be 001, meaning that we have to traverse left twice from the root (0 and 1 mean traverse left and right, respectively.) and finally insert the new edge block as the right child of the edge block that is left twice from the root. The process repeats for each new edge block of that adjacency. Then using the last insert edge offset and last insert edge block values stored in the edge sentinel node, edge insertions start from the empty spaces in the last edge block in the previous batch insertion. Once the edge block becomes full, inserts commence on the newly inserted edge blocks until all the edges get inserted, followed by the updates on the edge queue.

\begin{algorithm}
\DontPrintSemicolon
\KwIn{(a) Vertex dictionary ; (b) Insert edge batch in CSR format;  (c) Prefix sum vector of edge blocks for the batch; }
\KwOut{Graph data structure has insert batch edges.}
\For{each source vertex $v_i \in  \mathbf{G}$} 
{
//Create adjacency for the first batch to this vertex\;
\If{adjacency is empty }{
Create CBT structure for the adjacency with edge blocks\;
insert offset $\leftarrow$ 0\;
insert block $\leftarrow$ root of CBT\;
}
//Adjacency already exists from a previous batch update\;
\Else {
\For{each edge block allocated for the adjacency $e_j$}
{
Generate bit string for the edge block\;
Insert the edge block to the adjacency using the bit string\;
}
space remaining $\leftarrow$ edge block size - last insert edge offset\;

\If{space remaining}{

insert offset $\leftarrow$ last insert edge offset\;
insert block $\leftarrow$ last insert edge block\;

}
\Else {

insert offset $\leftarrow$ 0\;
insert block $\leftarrow$ $e_0$\;

}

}

start index $\leftarrow$ csr offset [$i$]\;
end index $\leftarrow$ csr offset [$i+1$]\;

\For{each edge $e_j$ between start index and end index in csr edges vector} {

Add edge $e_j$ to the edge block\;
Increment active edge count in edge block\;

\If{active edge count $\geq$ edge block size} {

Increment active edge count of adjacency with the number of edges added in insert block\;
Increment edge block count of adjacency\;
insert offset $\leftarrow$ 0\;
// Edge blocks of an adjacency during a batch are contiguous\;
insert block $\leftarrow$ insert block + 1\;

}

}

}
Update edge queue\;
\caption{Batched Edge Insertion}
\label{edge_algorithm}
\end{algorithm}


We also explored the possibility of parallelizing edge block insertions to the CBT within an adjacency. In this case, the number of threads launched by the GPU kernel would be equal to the number of edge blocks required by all the vertices in the current update batch. Each thread now would be responsible for appropriately assigning its edge block's pointers to make the CBT structure intact. Also, each thread only needs to fill the edge entries allocated to it. This procedure meant that there needed to be a lot more bookkeeping that needed to be done in the batch pre-processing stage, and it also delivered worse performance than the approach that used one thread per adjacency. The performance loss was due to the additional work each thread needed to do to keep the CBT structure intact.

\subsubsection{Dynamic edge deletes}

During dynamic edge deletes, the graph data structure assumes the update batch to be in a \textit{CSR} format as in the edge inserts case. The number of threads launched by the GPU delete kernel equals the number of vertices in the graph. Each thread then performs an in-order traversal over the edge blocks of the CBT starting from the root and, during the traversal, checks the edge entries within the edge block currently being traversed. If the edge entries match the deleted entries in the update batch, they are marked as deleted. Simply marking the edge entries as deleted creates holes in the edge blocks of the CBT as time progresses and could increase the memory footprint. We plan to devise proper compaction and memory reclamation mechanisms to address this issue in future.

\begin{algorithm}
\DontPrintSemicolon
\KwIn{(a) Vertex dictionary ; (b) Delete edge batch in CSR format; }
\KwOut{Updated graph data structure.}
\For{each source vertex $v_i \in  \mathbf{G}$} 
{
//Start and End index of edges to be deleted\;
start index $\leftarrow$ csr offset [$i$]\;
end index $\leftarrow$ csr offset [$i+1$]\;

Perform in-order traversal of the adjacency\;
\For{each edge block $EB_j$ encountered during traversal} {

\For{each edge $e_k$ in the edge block} {

\For{each edge $e_l$ between start and end index in CSR edges } {

\If{$e_k$ == $e_l$} {

Mark $e_k$ as deleted\;
Decrement active edge count of edge block\;
Decrement active edge count of adjacency\;

}

}

}





}

}
\caption{Batched Edge Deletes}
\label{delete_algorithm}
\end{algorithm}

\subsection{Querying operations}

The proposed work supports fast search queries for edges stored in the graph data structure. Given a set of source and destination vertex pairs, a search pre-processing GPU kernel gets launched with one thread that performs an in-order traversal of the CBT associated with the source vertex. During the traversal, the address of all the edge blocks in the CBT gets pushed to a vector. Later, another GPU kernel gets launched with the number of threads equal to the degree of the source vertex. The idea is that each thread searches only one edge entry in one edge block of the CBT. Once in the kernel, each thread depending on its unique identifier, can pinpoint the edge block and the edge entry within that edge block it needs to search. If one thread finds the searched edge, it returns true to the host. The search query timings are very low since one thread gets launched for each edge entry.

\begin{algorithm}
\DontPrintSemicolon
\KwIn{(a) Vertex dictionary ; (b) Source and destination vertices of the edge needed to be searched; }
\KwOut{Sets the device search flag as 1 if the edge is present, else 0.}
//One thread performs in-order traversal\;
Perform in-order traversal on the adjacency of the source vertex\;
\For{each edge block $EB_i$ encountered during traversal} {


search blocks[$i$] $\leftarrow$ Address of $EB_i$\;

}
//In-order traversal thread ends\;

//Threads launched from now is $addresses$ $pushed$ $*$ $edge block size$\;
\For{each thread $T_i$} 
{

edge block address $\leftarrow$ search blocks[$i$ / $edge block size$]\;
edge block offset $\leftarrow$ search blocks[$i$ \% $edge block size$]\;

extracted value $\leftarrow$ The destination entry on edge block address and edge block offset\;

\If{$extracted value$ == $destination$} {
device search flag $\leftarrow$ 1\;
}

}
//All threads terminated\;
\caption{Querying Operations on Edges}
\label{search_algorithm}
\end{algorithm}


\section{Performance Evaluation}
\label{perf_sec}
In this section, we compare the performance of the proposed method to existing state-of-the-art graphs data structures like \textit{Hornet} and \textit{faimGraph}. We have omitted \textit{cuSTINGER}, \textit{aimGraph}, \textit{GPMA}, and \textit{diff-CSR} from the comparison as they are considered obsolete and fell far behind \textit{Hornet} and \textit{faimGraph} in terms of batched edge insert and delete timings, as well as performance in real-world graph processing algorithms like BFS, SSSP, and Page Rank.


\begin{table}[tbh]
\centering    
\footnotesize{
\begin{tabular}{ | l | l | c | c | c | c | c |} 
 \hline
 S. No & Input Graph & |V| & |E| & $D_{max}$ & $D_{avg}$ & Size \\ [1ex] 
 \hline\hline
 G1 & co-papers-dblp & 540K & 30M & 3K & 56 & 200MB\\ 
 \hline 
 G2 & co-papers-citeseer & 434K & 32M & 1K & 73 & 207MB\\ 
 \hline 
 G3 & hugetrace-00020 & 16M & 48M & 3 & 2 & 398MB\\ 
 \hline
 G4 & channel-500x100x100-b050 & 5M & 86M & 18 & 17 & 663MB\\ 
 \hline
 G5 & delaunay\_n24 & 17M & 100M & 26 & 5 & 839MB\\
 \hline
 G6 & inf-europe\_osm & 51M & 108M & 13 & 2 & 949MB\\
 \hline
 G7 & rgg\_n\_2\_24\_s0 & 17M & 266M & 40 & 15 & 2.2GB\\ [1.5ex] 
 \hline

\end{tabular}
}
\caption{Properties of the input graphs taken for comparison. The columns represent the input graph name, the number of vertices, the number of edges, the maximum degree, the average degree, and the size of the input graph file, respectively.}
\end{table}


We compare the batched vertex and edge insert and delete timings for the performance evaluation. Later, we compare the memory consumption of \textit{Hornet}, \textit{faimGraph}, and the proposed work. Table 1 lists the properties of the input graphs taken for comparison. All the graphs are part of the 10th DIMACS implementation challenge. The experimental setup included an NVIDIA RTX 2060 12GB GPU with an Intel Core i3-12100F CPU and 16GB DDR4 3200MHz RAM. The GPU has 2,176 CUDA cores and 34 SMs (Streaming Multi-processors).

\subsection{Initialization}

The initialization phase involves the necessary preprocessing required for the input graph. The edge block size, which determines the number of edge entries an edge block can hold, is set as the average degree of the non-zero degree vertices in the first batch. Experimental results have shown that the average degree of the non-zero degree vertices stays along the same lines over future batches. The vertex dictionary gets initialized with a size equal to the closest power of two, and edge sentinel nodes are attached to each source vertex in the vertex dictionary. Then one \textit{cudaMalloc} call reserves half the GPU memory for edge blocks and pushes the addresses into the edge queue.

\begin{center}

\begin{tabular}{ | l | c | c | c |} 
 \hline
 Input & Hornet & faimGraph & GraphVine \\ [1ex] 
 \hline\hline
 G1 & 643.973ms & 16.981ms & 0.575ms\\ 
 \hline 
 G2 & 538.074ms & 19.915ms & 0.589ms\\ 
 \hline 
  G3 & 15617ms & 43.831ms & 2.227ms\\ 
 \hline
 G4 & 4929ms & 50.772ms & 1.355ms\\ 
 \hline
 G5 & 16059ms & 62.199ms & 2.386ms\\
 \hline
 G6 & 51298ms & 119.028ms & 4.042ms\\
 \hline
 G7 & 17552ms & 140.11ms & 3.180ms\\ 
 \hline
\end{tabular}
\end{center}
\textbf{Table 2: Comparison of the initialization timings of the different graph data structures. The timings are in milliseconds(ms).}

\textit{Figure 5} shows the initialization timings for the different graph data structures across the input graphs. Experiments showed that the initialization timings for the graph data structures for different batch update sizes remained the same. The proposed work significantly improved the initialization timings, where it gave 111895\%, 91253\%, 701157\%, 363663\%, 672951\%, 1269024\%, and 551849\% improvements over \textit{Hornet} and 2853\%, 3281\%, 1968\%, 3647\%, 2506\%, 2844\%, and 4306\% improvements over \textit{faimGraph} on \textit{G1, G2, G3, G4, G5, G6, and G7}, respectively.

We attribute this significant improvement in initialization timings to the very few \textit{cudaMalloc} calls we perform (3 calls) to reserve GPU memory, namely, for the vertex dictionary, the edge sentinel nodes, and the edge blocks.

\begin{center}
\begin{figure}
\includegraphics[width=0.8\columnwidth]{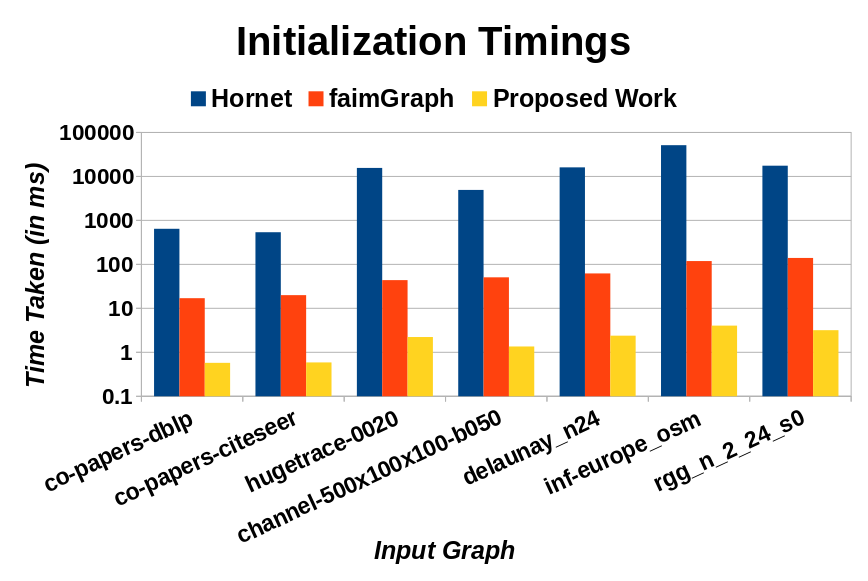} \\
\caption{Initialization timings for the different graph data structures across different input graphs. Please note that the y-axis of the graph is in a logarithm scale.}
\end{figure}
\end{center}

\subsection{Batched edge insert timings}

The batched edge inserts involve inserting an edge batch into the graph data structure. Once the graph data structure gets initialized with the vertex dictionary and the edge sentinel nodes, the edge batch update in \textit{CSR} format serves as input to the data structure. The prefix sum vector for edge blocks gets calculated for the parallel popping of edge blocks from the edge queue. Then the batch update is copied to the GPU, followed by a GPU kernel call for edge insertion. We conducted experiments on different batch sizes for the graph data structures across the different graphs, followed by a bulk build of the data structure where all the edges in the input graph get inserted in one go.

\textit{Figures 6, 7, 8, and 9} show the batched edge insert timings for different batch sizes. For a batch size of 100K edges, the proposed work, when compared to \textit{Hornet}, showed improvements of 1407\%, 1227\%, 492\%, 809\%, and 290\% for\textit{ G1, G2, G5, G6, and G7}, respectively in edge insert timings. However, \textit{Hornet} performed better in \textit{G3 and G4} by 107\% and 16\%, respectively. Compared to \textit{faimGraph} with the same batch size of 100K, the proposed work did give improved timings by 41\%, 74\%, and 234\% in \textit{G1, G2, and G5}, respectively. \textit{faimGraph} performed better by 205\%, 111\%, 585\%, and 71\% in \textit{G3, G4, G6, and G7}, respectively compared to the proposed work.

When the edge insert batch size increased to 1M edges, the proposed work gave improvements over \textit{Hornet} by 3611\%, 3384\%, 510\%, 57\%, 3322\%, 6813\%, and 1427\% for \textit{G1, G2, G3, G4, G5, G6, and G7}, respectively. Compared to \textit{faimGraph}, the proposed work gave improvements of 305\%, 357\%, 30\%, 145\%, 47\%, and 6\% for \textit{G1, G2, G3, G4, G5, and G7,} respectively. \textit{faimGraph}, however, performed better than the proposed work on\textit{ G6} by 58\%.

\begin{center}
\begin{figure}
\includegraphics[width=0.8\columnwidth]{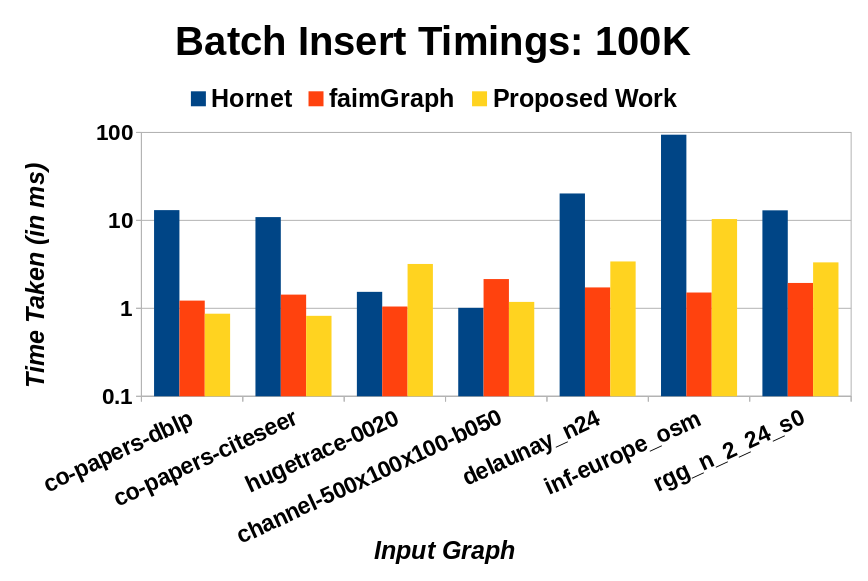} \\
\caption{Batch insert timings for a batch size of 100K for the different graph data structures across different input graphs.}
\end{figure}
\end{center}

\begin{center}
\begin{figure}
\includegraphics[width=0.8\columnwidth]{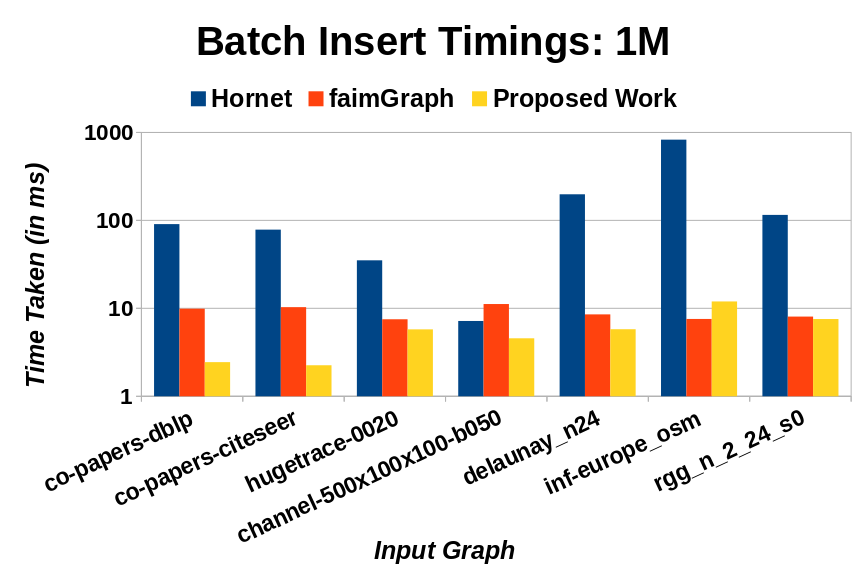} \\
\caption{Batch insert timings for a batch size of 1M for the different graph data structures across different input graphs.}
\end{figure}
\end{center}

For the edge insert batch size of 10M, the proposed work performed better than \textit{Hornet} by 864\%, 709\%, 7133\%, 61\%, 8115\%, 30047\%, and 2749\% on \textit{G1, G2, G3, G4, G5, G6, and G7,} respectively. \textit{faimGraph} has a maximum batch size cap of 1M, so we could not run \textit{faimGraph} with a batch size of 10M.

\begin{center}
\begin{figure}
\includegraphics[width=0.8\columnwidth]{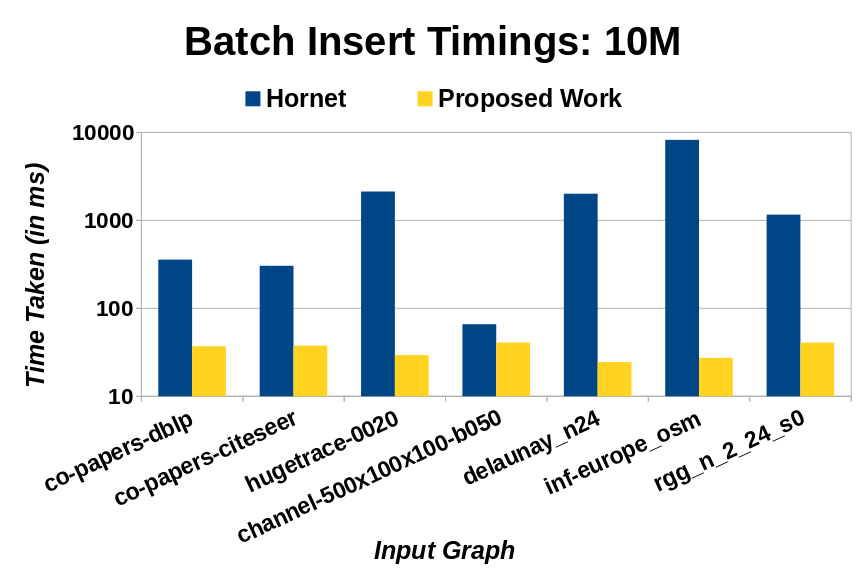} \\
\caption{Batch insert timings for a batch size of 10M for the different graph data structures across different input graphs. \textit{faimGraph} is absent in this comparison since it doesn't support batches in size more than 1M.}
\end{figure}
\end{center}

When it came to bulk builds, where the entire input graph gets inserted into the graph data structure in one go, the proposed work gave improvements of 326\%, 235\%, 8592\%, 1068\%, and 5450\% on  \textit{G1, G2, G3, G4, and G5}, respectively. The graphs \textit{G6 and G7} ran out of memory and crashed on \textit{Hornet}, while the proposed work had no trouble handling it. \textit{faimGraph} could not perform bulk build due to their maximum batch size cap of 1M edges.

\begin{center}
\begin{figure}
\includegraphics[width=0.8\columnwidth]{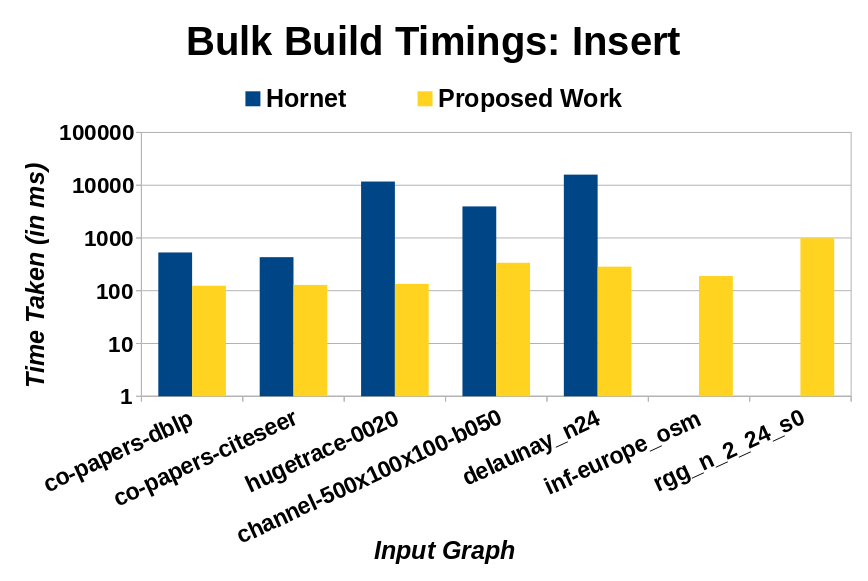} \\
\caption{Bulk build insert timings for the different graph data structures across different input graphs. \textit{faimGraph} is absent in this comparison since it doesn't support batches in size more than 1M. \textit{Hornet} ran out of memory and crashed in the last two graphs.}
\end{figure}
\end{center}

{\footnotesize
\begin{center}
\begin{tabular}{ | l | c | c | c | c |} 
 \hline
 Input & Batch Size & Hornet & faimGraph & GraphVine \\ [1ex] 
 \hline\hline
 G1 & 100K & 13.083ms & 1.224ms & 0.868ms\\ 
 & 1M & 90.699ms & 9.909ms & 2.444ms\\ 
 & 10M & 358.091ms & NA & 37.11ms\\ 
  & Bulk & 531.632ms & NA & 124.647ms\\ 
 \hline 
 G2 & 100K & 10.9ms & 1.431ms & 0.821ms\\ 
    
  & 1M & 78.501ms & 10.312ms & 2.253ms\\ 
& 10M & 304.825ms & NA & 37.658ms\\ 
& Bulk & 432.848ms & NA & 129.029ms\\ 
 \hline 
  G3 & 100K & 1.539ms & 1.048ms & 3.197ms\\ 

  & 1M & 35.181ms & 7.504ms & 5.762ms\\ 
& 10M & 2129.69ms & NA & 29.444ms\\ 
& Bulk & 11747.7ms & NA & 135.142ms\\ 
 \hline
 G4 & 100K & 1.013ms & 2.153ms & 1.183ms\\ 

  & 1M & 7.179ms & 11.184ms & 4.563ms\\ 
& 10M & 65.983ms & NA & 40.887ms\\ 
& Bulk & 3967.16ms & NA & 339.383ms\\ 
 \hline
 G5 & 100K & 20.245ms & 1.730ms & 3.415ms\\ 

  & 1M & 198.051ms & 8.531ms & 5.787ms\\ 
& 10M & 2011.45ms & NA & 24.485ms\\ 
& Bulk & 15878.7ms & NA & 286.08ms\\ 
 \hline
 G6 & 100K & 94.263ms & 1.512ms & 10.362ms\\ 

  & 1M & 827.766ms & 7.572ms & 11.973ms\\ 
& 10M & 8235.38ms & NA & 27.317ms\\ 
& Bulk & NA & NA & 190.068ms\\ 
 \hline
 G7 & 100K & 13.02ms & 1.945ms & 3.333ms\\ 

  & 1M & 115.439ms & 8.058ms & 7.559ms\\ 
& 10M & 1161.95ms & NA & 40.779ms\\ 
& Bulk & NA & NA & 992.54ms\\ 
 \hline
\end{tabular}
\end{center}}
\textbf{Table 3: Comparison of the batched edge insert timings of the different graph data structures across different batch sizes. The timings are in milliseconds(ms).}

The experimental results suggest that the performance of the proposed method increases drastically as the batch size increases, attributed to more time spent in edge insertions compared to fixing the CBT structure during an update for large batches. For a fixed batch size, the proposed work gave the best performance when the average and maximum degrees of the graph were either very far apart or very close. If the average and maximum degrees were close, the height of the CBT of adjacencies would be less, meaning less time is spent fixing the CBT structure. If the average and maximum degrees were far apart, the time spent for CBT creation is amortized by the large number of edges inserted into that adjacency. The proposed work gave the worst results when the number of vertices in the input graph was high but still managed to perform better than \textit{Hornet} and come close to \textit{faimGraph} as the batch sizes increased.

\subsection{Batched edge delete timings}

The batched edge deletes involve deleting a group of edges inserted into the graph data structure in a previous batch update. The delete edge batch update in \textit{CSR} format serves as input to the data structure, similar to edge insertion. Similar to inserts, experiments were conducted on different batch sizes for the graph data structures across the different graphs, followed by a bulk build of the data structure.

\begin{center}
\begin{figure}
\includegraphics[width=0.8\columnwidth]{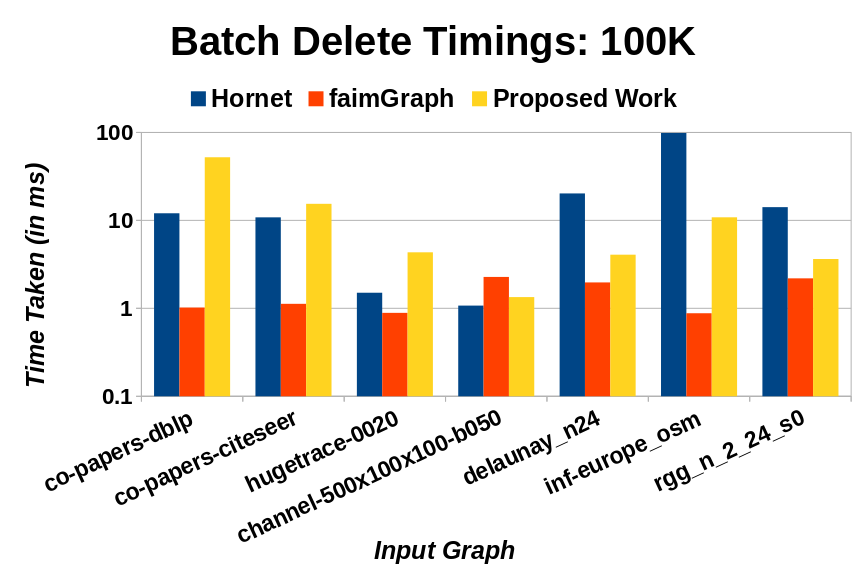} \\
\caption{Batch delete timings for a batch size of 100K for the different graph data structures across different input graphs.}
\end{figure}
\end{center}

\textit{Figures 10, 11, 12, and 13} show the batched edge delete timings for the different batch sizes. For a batch size of 100K edges, the proposed work gave improvements over \textit{Hornet} by 596\%, 810\%, and 288\% on G5, G6, and G7, respectively. \textit{Hornet} had better timings over the proposed work by 333\%, 42\%, 188\%, and 24\% on \textit{G1, G2, G3, and G4}, respectively. Compared to \textit{faimGraph}, the proposed work gave better timings by 70\% on \textit{G4}. \textit{faimGraph} gave better timings over the proposed work by 14770\%, 7068\%, 388\%, 106\%, 1132\%, and 66\% on \textit{G1, G2, G3, G5, G6, and G7}, respectively.

\begin{center}
\begin{figure}
\includegraphics[width=0.8\columnwidth]{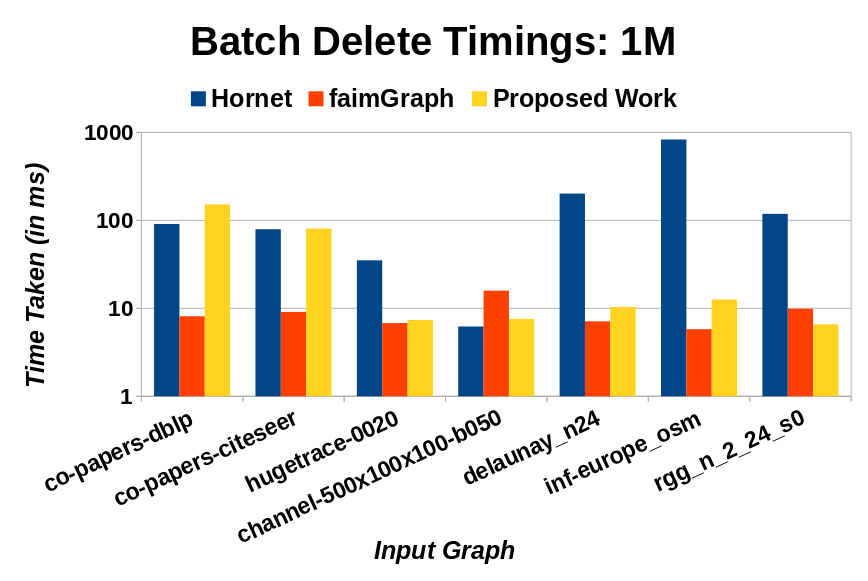} \\
\caption{Batch delete timings for a batch size of 1M for the different graph data structures across different input graphs.}
\end{figure}
\end{center}

The edge delete batch size of 1M edges gave the proposed work improvements over \textit{Hornet} by 377\%, 1847\%, 6494\%, and 1707\% on \textit{G3, G5, G6, and G7}, respectively. The input graph \textit{G2} had a tie, while \textit{Hornet} outperformed the proposed work by 66\% and 21\% on \textit{G1 and G4}, respectively. Compared to \textit{faimGraph}, the proposed work improved timings by 110\% and 50\% on \textit{G4 and G7}, respectively. The proposed work lost out to \textit{faimGraph} by 7652\%, 785\%, 8\%, 511\%, and 117\% on \textit{G1, G2, G3, G5, and G6}, respectively.

\begin{center}
\begin{figure}
\includegraphics[width=0.8\columnwidth]{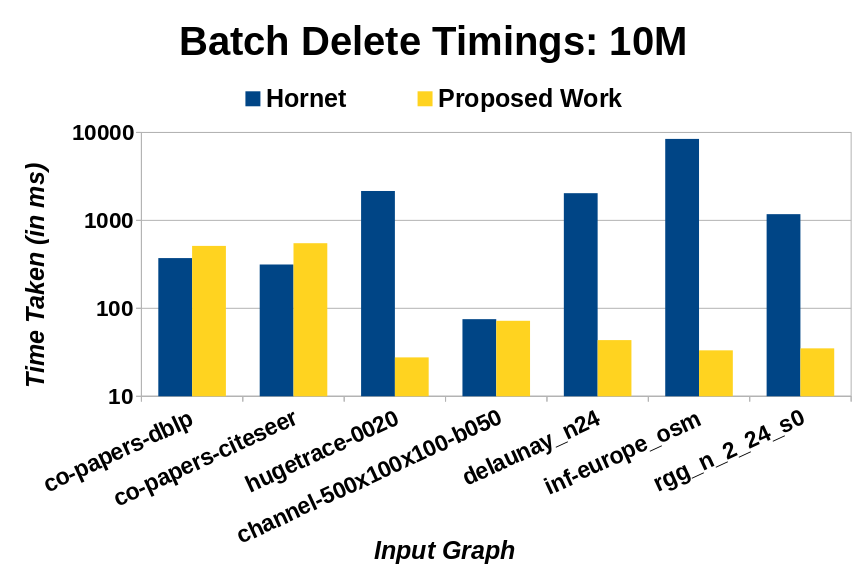} \\
\caption{Batch delete timings for a batch size of 10M for the different graph data structures across different input graphs. \textit{faimGraph} is absent in this comparison since it doesn't support batches in size more than 1M.}
\end{figure}
\end{center}

\begin{center}
\begin{figure}
\includegraphics[width=0.8\columnwidth]{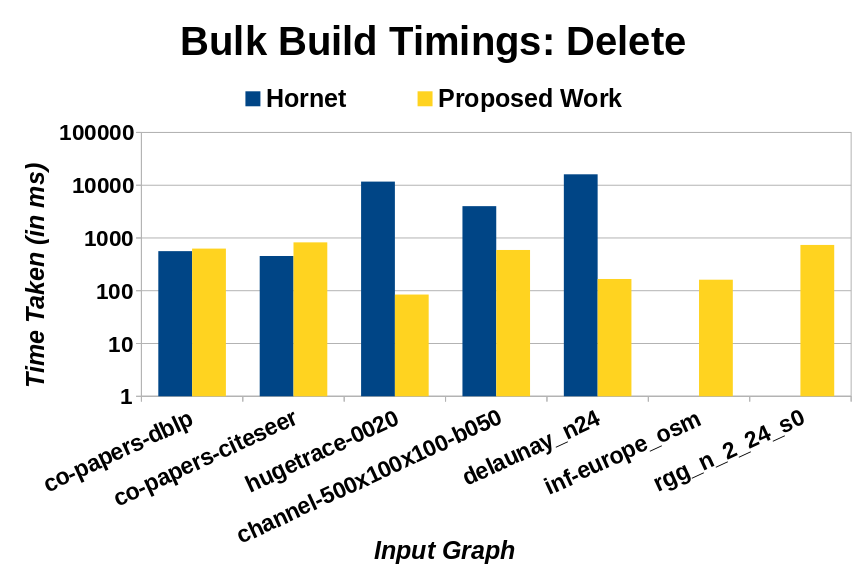} \\
\caption{Bulk build delete timings for the different graph data structures across different input graphs. \textit{faimGraph} is absent in this comparison since it doesn't support batches in size more than 1M. \textit{Hornet} ran out of memory and crashed in the last two graphs.}
\end{figure}
\end{center}

When the edge deletes batch size increased to 10M edges, the proposed work gave improved timings by 7706\%, 4\%, 4585\%, 25262\%, and 3257\% over \textit{Hornet} on \textit{G3, G4, G5, G6, and G7}, respectively. \textit{Hornet} gave improved timings by 37\% and 74\% on \textit{G1 and G2}, respectively. Like before, \textit{faimGraph} could not process batches of size more than 1M.

Bulk edge delete batches gave the proposed work improvements over \textit{Hornet} by 13717\%, 576\%, and 9537\% on\textit{ G3, G4 and G5}, respectively. \textit{Hornet} had improved timings over the proposed work by 11\% and 81\% on \textit{G1 and G2}, respectively. The input graphs \textit{G6 and G7} ran out of memory and crashed on \textit{Hornet}, while the proposed work ran them flawlessly.

Observations from the experiments suggest that the proposed work gave the best results when the average degree of the graph was close to its maximum since, in such cases, the height of an adjacency's CBT would be the least. The proposed work gave poor performance when there was a stark difference in the average and maximum degrees, attributed to the increased height of the CBT of each adjacency and hence the additional time required to traverse the entire CBT. Another observation was that the performance impact is low if only a few such adjacencies exist, and issues arise when most adjacencies have a significant height.

{\footnotesize
\begin{center}
\begin{tabular}{ | l | c | c | c | c |} 
 \hline
 Input & Batch Size & Hornet & faimGraph & GraphVine \\ [1ex] 
 \hline\hline
 G1 & 100K & 12.046ms & 1.020ms & 52.207ms\\ 
 & 1M & 91.146 & 8.115ms & 151.679ms\\ 
 & 10M & 372.862ms & NA & 513.151ms\\ 
  & Bulk & 563.705ms & NA & 629.137ms\\ 
 \hline 
 G2 & 100K & 10.832ms & 1.125ms & 15.432ms\\ 
    
  & 1M & 79.254ms & 9.108ms & 80.649ms\\ 
& 10M & 315.28ms & NA & 550.459ms\\ 
& Bulk & 455.584ms & NA & 827.061ms\\ 
 \hline 
  G3 & 100K & 1.504ms & 0.888ms & 4.336ms\\ 

  & 1M & 35.143ms & 6.801ms & 7.367ms\\ 
& 10M & 2159.95ms & NA & 27.670ms\\ 
& Bulk & 11706.7ms & NA & 84.723ms\\ 
 \hline
 G4 & 100K & 1.074ms & 2.276ms & 1.342ms\\ 

  & 1M & 6.214ms & 15.871ms & 7.559ms\\ 
& 10M & 75.283ms & NA & 72.130ms\\ 
& Bulk & 4011.89ms & NA & 593.154ms\\ 
 \hline
 G5 & 100K & 20.25ms & 1.972ms & 4.077ms\\ 

  & 1M & 201.747ms & 7.114ms & 10.359ms\\ 
& 10M & 2039.19ms & NA & 43.523ms\\ 
& Bulk & 16104.4ms & NA & 167.104ms\\ 
 \hline
 G6 & 100K & 98.674ms & 0.879ms & 10.837ms\\ 

  & 1M & 831.202ms & 5.791ms & 12.605ms\\ 
& 10M & 8442.63ms & NA & 33.288ms\\ 
& Bulk & NA & NA & 161.929ms\\ 
 \hline
 G7 & 100K & 14.144ms & 2.194ms & 3.641ms\\ 

  & 1M & 118.606ms & 9.885ms & 6.563ms\\ 
& 10M & 1177.21ms & NA & 35.059ms\\ 
& Bulk & NA & NA & 738.115ms\\ 
 \hline
\end{tabular}
\end{center}}
\textbf{Table 4: Comparison of the batched edge delete timings of the different graph data structures across different batch sizes. The timings are in milliseconds(ms).}

\subsection{Memory consumption}

An essential aspect of graph data structure is memory consumption since it determines the size of the input graphs that the data structure can handle. If the graph data structure consumes too much memory, the program will crash on large graphs, limiting its capability. Here we compare the memory footprint of the graph data structures across the different graphs. Using the inbuilt cudaGetMemInfo function, we calculate the GPU memory usage before the graph data structure launch and after edge insertions for different batch sizes, followed by a bulk build.

\begin{center}
\begin{figure}
\includegraphics[width=0.8\columnwidth]{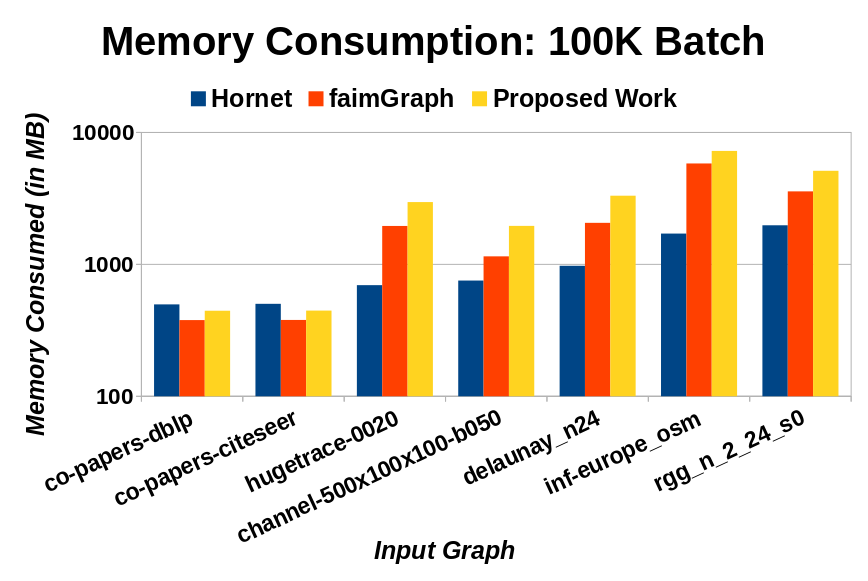} \\
\caption{Memory consumption for the batch size of 100K for the different graph data structures across different input graphs.}
\end{figure}
\end{center}

\textit{Figures 14, 15, 16 and 17} show the memory consumption for the different batch sizes. For a batch size of 100K, the proposed work consumed 12\% and 13\% less memory than \textit{Hornet} on \textit{G1 and G2}. However, the proposed work consumed more memory than \textit{Hornet} by 326\%, 159\%, 239\%, 323\%, and 158\% on \textit{G3, G4, G5, G6 and G7}, respectively. Compared to \textit{faimGraph}, the proposed work consumed 20\%, 17\%, 51\%, 70\%, 60\%, 24\%, and 43\% more memory on \textit{G1}, \textit{G2}, \textit{G3, G4, G5, G6 and G7}, respectively.

\begin{center}
\begin{figure}
\includegraphics[width=0.8\columnwidth]{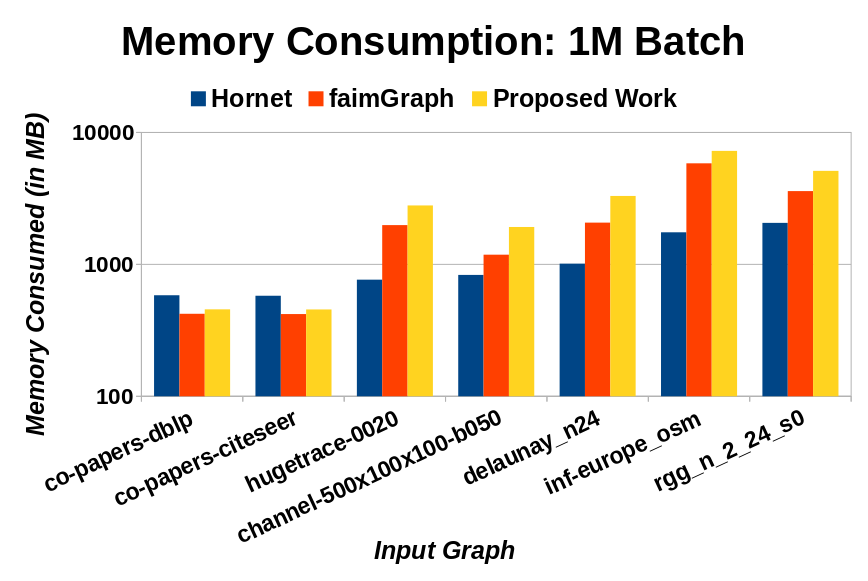} \\
\caption{Memory consumption for the batch size of 1M for the different graph data structures across different input graphs.}
\end{figure}
\end{center}

For a batch size of 1M, the proposed work consumed 27\% less memory than \textit{Hornet} for both \textit{G1 and G2}, but consumed 265\%, 130\%, 226\%, 313\%, and 147\% more memory on \textit{G3, G4, G5, G6 and G7}, respectively. Compared to \textit{faimGraph}, the proposed work consumed 8\%, 8\%, 40\%, 73\%, 59\%, and 42\% more memory on \textit{G1, G2, G3, G4, G5, and G7}, respectively. Nevertheless, the proposed work consumed 24\% less memory than \textit{faimGraph} on \textit{G6}.

\begin{center}
\begin{figure}
\includegraphics[width=0.8\columnwidth]{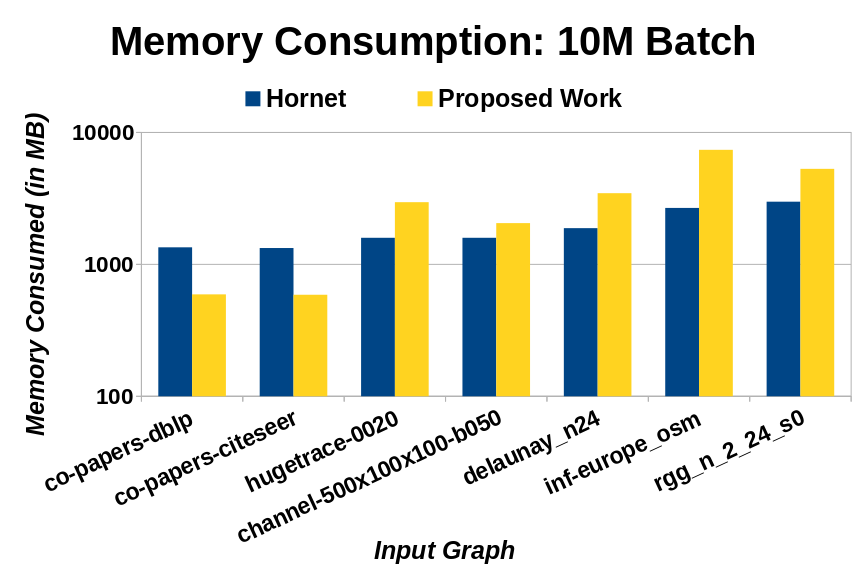} \\
\caption{Memory consumption for the batch size of 10M for the different graph data structures across different input graphs. \textit{faimGraph} is absent in this comparison since it doesn't support batches in size more than 1M.}
\end{figure}
\end{center}

\begin{center}
\begin{figure}
\includegraphics[width=0.8\columnwidth]{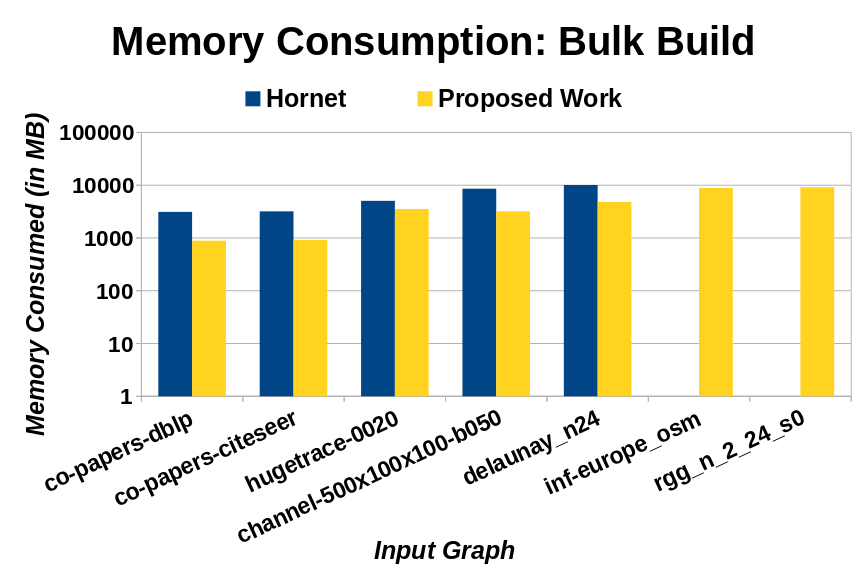} \\
\caption{Memory consumption on bulk build for the different graph data structures across different input graphs.\textit{faimGraph} is absent in this comparison since it doesn't support batches in size more than 1M. \textit{Hornet} ran out of memory and crashed in the last two graphs.}
\end{figure}
\end{center}


For batch sizes of 10M, the proposed work consumed 128\% and 126\% less memory than \textit{Hornet} on\textit{ G1 and G2}, respectively. However, \textit{Hornet} consumed less memory than the proposed work by 86\%, 29\%, 84\%, 175\%, and 77\% on \textit{G3, G4, G5, G6, and G7}, respectively. We have no data on \textit{faimGraph} for this batch size since it does not support batch updates larger than 1M.

For bulk builds, the proposed work got on top by consuming 250\%, 247\%, 42\%, 168\%, and 109\% less memory than \textit{Hornet} on \textit{G1, G2, G3, G4, and G5}, respectively. The input graphs \textit{G6 and G7} ran out of memory and crashed on \textit{Hornet}, while the proposed work ran without any issues and consumed memory close to the 10M batch size.

{\footnotesize
\begin{center}
\begin{tabular}{ | l | c | c | c | c |} 
 \hline
 Input & Batch Size & Hornet & faimGraph & GraphVine \\ [1ex] 
 \hline\hline
 G1 & 100K & 497MB & 378MB & 445MB\\ 
 & 1M & 583MB & 422MB & 456MB\\ 
 & 10M & 1347MB & NA & 592MB\\ 
  & Bulk & 3124MB & NA & 890MB\\ 
 \hline 
 G2 & 100K & 502MB & 379MB & 446MB\\ 
    
  & 1M & 578MB & 420MB & 455MB\\ 
& 10M & 1330MB & NA & 588MB\\ 
& Bulk & 3197MB & NA & 921MB\\ 
 \hline 
  G3 & 100K & 696MB & 1955MB & 2966MB\\ 

  & 1M & 766MB & 1985MB & 2797MB\\ 
& 10M & 1590MB & NA & 2961MB\\ 
& Bulk & 5059MB & NA & 3550MB\\ 
 \hline
 G4 & 100K & 754MB & 1150MB & 1958MB\\ 

  & 1M & 832MB & 1184MB & 1921MB\\ 
& 10M & 1590MB & NA & 2055MB\\ 
& Bulk & 8590MB & NA & 3201MB\\ 
 \hline
 G5 & 100K & 976MB & 2065MB & 3315MB\\ 

  & 1M & 1013MB & 2072MB & 3302MB\\ 
& 10M & 1882MB & NA & 3465MB\\ 
& Bulk & 10102MB & NA & 4831MB\\ 
 \hline
 G6 & 100K & 1712MB & 5821MB & 7245MB\\ 

  & 1M & 1751MB & 5834MB & 7247MB\\ 
& 10M & 2679MB & NA & 7385MB\\ 
& Bulk & NA & NA & 8908MB\\ 
 \hline
 G7 & 100K & 1979MB & 3578MB & 5120MB\\ 

  & 1M & 2064MB & 3592MB & 5115MB\\ 
& 10M & 2986MB & NA & 5297MB\\ 
& Bulk & NA & NA & 9142MB\\ 
 \hline
\end{tabular}
\end{center}}
\textbf{Table 5: Comparison of the memory consumption of the different graph data structures across different batch sizes. Memory usage is denoted in MegaBytes(MB).}

An observation made was that the memory consumption of both the proposed work and \textit{faimGraph} increased drastically when the number of vertices in the graph got large. For the proposed work, more vertices meant more edge sentinel nodes, which meant higher memory usage. The memory usage of all the data structures increased with an increase in batch size. However, \textit{Hornet} had an exponential increase in memory usage for large batches compared to small ones, despite consuming significantly less memory than the proposed work and \textit{Hornet} for vertices with a large number of vertices on small batches. The increase in memory consumption for \textit{Hornet} was so significant that it ran out of memory for bulk builds on \textit{G6 and G7}. The proposed work consumed more memory than the competition for small batches, had a linear increase in memory consumption as the batch size got bigger, and had no issues running \textit{G6 and G7} while having 3GB of GPU memory still available.


\section{Conclusion \& Future Scope}
\label{conclusion_sec}

The proposed graph data structure performs fast insert, delete, and querying performance on dynamic graphs and significantly outperforms the existing state-of-the-art graph data structures on large batch sizes. We recorded over 300000\% and 2500\% improvement in initialization timings over \textit{Hornet} and \textit{faimGraph}, respectively, across all graphs. For batched edge insertions, we observed over than 1000\% improvement over \textit{Hornet} and greater than 150\% improvement over \textit{faimGraph} in multiple instances. However, in small batch sizes, the proposed work gave worse timings than the competition for some graphs. The batched edge deletes followed a similar pattern, where over 5000\% improvements were detected over \textit{Hornet} for large batch sizes. In contrast, \textit{faimGraph} outperformed both \textit{Hornet} and the proposed work for small batch sizes. The proposed work also consumed far less memory than the competition as the batch sizes got larger.

There is still scope for improving the batched edge insertion times by splitting the GPU kernel responsible for insertions into two, where the first kernel fixes the CBT structure. In contrast, the second kernel performs edge insertions, allowing us to launch more threads with the second kernel, which could reduce insert timings. The memory footprint can also be reduced by dynamically changing the edge block size, which could reduce internal fragmentation within edge blocks. This is in line with the fact that over future batch updates, attributes of the graph, like the average degree of vertices across batches, could change. We also plan to optimize the batched edge deletes further and integrate the proposed work with real-world graph algorithms like Breadth-First-Search, Single-Source-Shortest-Path, and Dijkstra's shortest path.

\begin{acks}                            
  This material is based upon work supported by the
  \grantsponsor{GS100000001}{National Science
    Foundation}{http://dx.doi.org/10.13039/100000001} under Grant
  No.~\grantnum{GS100000001}{nnnnnnn} and Grant
  No.~\grantnum{GS100000001}{mmmmmmm}.  Any opinions, findings, and
  conclusions or recommendations expressed in this material are those
  of the author and do not necessarily reflect the views of the
  National Science Foundation.
\end{acks}



\bibliography{bibliography}
\end{document}